\documentclass[aps,prl, twocolumn, superscriptaddress]{revtex4-1}
\usepackage{graphicx}
\usepackage{amsmath}
\usepackage{amsfonts}
\usepackage{amsthm}
\usepackage{amssymb}
\usepackage{amsbsy}
\usepackage{wasysym}
\usepackage{bm}
\usepackage{mathrsfs}
\usepackage{color}
\usepackage{hyperref}

\usepackage{pdfpages}

\newcommand{\beq}{\begin{equation}}
\newcommand{\eeq}{\end{equation}}
\newcommand{\bea}{\begin{eqnarray}}
\newcommand{\eea}{\end{eqnarray}}

\newlength{\myL}

\newcommand{\e}{\varepsilon}

\renewcommand{\(}{\left(}
\renewcommand{\)}{\right)}

\def\ket#1{{\left|#1\right\rangle}}

\usepackage{ulem}

\def\be{\begin{eqnarray}}
\def\ee{\end{eqnarray}}

\begin{document}

\title{Quantum criticality of hot random spin chains}
\author {R. Vasseur}
\affiliation{Department of Physics, University of California, Berkeley, CA 94720, USA}
\affiliation{Materials Science Division, Lawrence Berkeley National Laboratories, Berkeley, CA 94720}
\author{A.C. Potter}
\affiliation{Department of Physics, University of California, Berkeley, CA 94720, USA}
\author{S. A. Parameswaran}
\affiliation{Department of Physics and Astronomy, University of California, Irvine, CA 92697, USA}

\date{\today}
\begin{abstract} 
{
We study the infinite-temperature properties of an infinite sequence of random quantum spin chains using a real-space renormalization group approach, and demonstrate that they exhibit non-ergodic behavior at strong disorder. The analysis is  conveniently implemented in terms of SU(2)$_k$ anyon chains that include the Ising and Potts chains as notable examples. Highly excited eigenstates of these systems exhibit properties 
usually associated with quantum critical ground states, leading us to dub them ``quantum critical glasses''. We argue that random-bond Heisenberg chains self-thermalize and that the excited-state entanglement crosses over from volume-law to logarithmic scaling at a length scale that diverges in the Heisenberg limit $k\rightarrow\infty$. The excited state fixed points are generically distinct from their ground state counterparts, and represent novel non-equilibrium critical phases of matter.
}

\end{abstract}
\maketitle

Quantum spin systems are central to condensed matter physics, underpinning aspects of the field as diverse as the theory of quantum critical phenomena~\cite{sachdev2011} to the role of topology~\cite{PhysRevLett.50.1153} and symmetry~\cite{PhysRevB.85.075125} in delineating zero-temperature phases of matter. Much is therefore known about their ground states and low-lying spectra. More recently,  spurred in part by the ability to experimentally probe such systems in the absence of external sources of equilibration~\cite{RevModPhys.80.885}, there has been growing interest in understanding whether they can thermalize in isolation~\cite{PhysRevA.43.2046,PhysRevE.50.888} --- an issue dictated by the behavior 
of excited states at {non-vanishing} energy density. 
Although most isolated quantum many-body systems `self-thermalize', acting as their own heat bath, a handful of {\it many-body localized} (MBL) systems --  typically one-dimensional systems with quenched randomness~\cite{Basko20061126,PhysRevB.75.155111,PhysRevB.82.174411,PhysRevLett.110.067204,2013arXiv1311.7151Y,2014arXiv1404.0686N,PhysRevX.4.011052,2014arXiv1403.1568K} -- instead exhibit non-ergodic dynamics and quantum glassiness. 
In contrast to self-thermalizing systems, whose excited states are highly entangled and classically incoherent, MBL systems exhibit robust quantum coherence analogous to {\it gapped} quantum ground-states --- including short-range (boundary-law) entanglement structure~\cite{BauerNayak} and exponentially decaying spatial correlations. These peculiar properties enable MBL systems to violate many standard tenets of equilibrium statistical mechanics --- raising the possibility of symmetry breaking and topological order~\cite{PhysRevB.88.014206,BauerNayak,PhysRevB.89.144201,2013arXiv1307.4092B} and quantum coherent dynamics at infinite effective temperature~\cite{2013arXiv1307.4092B}.

{In this Letter, we construct and study an infinite family of random quantum spin chains that, like MBL systems, exhibit non-ergodic quantum coherent dynamics characterized by the absence of thermal transport. However, unlike ordinary MBL systems, whose eigenstates behave like gapped equilibrium ground states, these models exhibit scale-free properties typical of zero-temperature gapless or quantum critical one-dimensional systems in  arbitrarily high energy excited states --- including
power-law decay of (disorder averaged) correlation functions and logarithmic scaling of the entanglement entropy of subsystems with their length.  We dub these phases ``quantum critical glasses" (QCGs), since, in addition to their scale-free critical properties, they exhibit slow glassy dynamics characterized by power law scaling of length $L$ and the {\it logarithm} of time,
 $t$: $\log t   \sim L^\psi$, where $0<\psi<1$ is a universal exponent characterizing the QCG phase. For example, tunneling through a length $L$ QCG chain takes characteristic time $\log t
  \sim L^\psi$; equivalently, the entanglement growth at time $t$ after a global quench grows as $S(t)\sim \log^{1/\psi} t $
  ~\cite{PhysRevLett.110.260601,2013arXiv1305.4915H,PhysRevLett.110.067204,PhysRevLett.112.217204}.}

{As we show, the scale-free 
nature of 
 QCGs enables an asymptotically exact computation of their universal scaling properties ({\it e.g.} the exponent $\psi$) of their dynamics on long length or time scales 
  via a real-space renormalization group (RSRG) approach~\cite{PhysRevLett.110.067204,PhysRevX.4.011052,PhysRevLett.112.217204}, providing a rare example of analytically exact results in interacting, disordered, and out-of-equilibrium systems.  Specifically, we consider random chains of anyonic spins~\cite{PhysRevLett.98.160409,Trebst01062008,Bonesteel,FidkowskiPRB08,FidkowskiPRB09}, described by a truncated version of the familiar SU(2) algebra, SU(2)$_k$, labeled by an integer $k$. The QCGs we study appear as renormalization group (RG) fixed points 
   in this family of models, and include the anyonic duals of  familiar spin chains such as the  random Ising ($k=2$) and Potts ($k=4$) models, as well as an infinite number of other  examples that correspond to random analogs of the minimal model and parafermionic conformal field theories (CFTs). These  exhaust many of the familiar 1D universality classes with potential QCG analogs. The formulation in terms of anyons is largely a matter of technical convenience, enabling us to simultaneously compute the properties for all $k$ on the same footing and elucidate the general scaling structure of a broad class of QCG fixed points.}

{We find that while QCGs share many common features with zero-temperature random critical points~\cite{FisherRSRG1,FisherRSRG2,PhysRevLett.75.4302,PhysRevB.55.12578,DamleHuse}, they generically have universal exponents that are distinct from those of their zero-temperature counterparts. Thus, with the (non-generic) exception of the previously studied Ising QCG~\cite{PhysRevX.4.011052}, they  represent new dynamical phases or critical points that emerge only in the out-of-equilibrium context.}

As noted above, our primary tool is a real-space renormalization group  (RSRG) procedure~\cite{PhysRevLett.43.1434,PhysRevB.22.1305,FisherRSRG1,FisherRSRG2,DamleHuse} that has been used to study the ground state properties of random-bond spin chains. The RSRG decimates couplings in a hierarchical fashion in which strong bonds are eliminated before weaker ones, either {`decimating' nearest-neighbor spins into singlets or `fusing' them into new effective superspins}. The effective disorder strength grows under this procedure, and so the resulting low-energy behavior is asymptotically exact and is governed by the so-called {\it infinite randomness fixed point}~\cite{FisherRSRG2}. Our strategy here 
is to adapt RSRG techniques to study the behavior of highly excited eigenstates and the resulting dynamics.

To this end, it is instructive to consider the single prior example of a QCG: the critical point of the random transverse-field Ising model. Ref.~\cite{PhysRevX.4.011052}  generalized the  ground state RSRG technique~\cite{FisherRSRG1}  to study the entire many-body spectrum by observing that at each step in the decimation procedure, it is possible to project into the excited-state manifold(s) rather than to the ground state. This yields a {choice of $M$ possible} decimations for each bond (for instance, $M=2$ for the random transverse-field Ising model~\cite{PhysRevX.4.011052}). Formally, following each choice leads to a rapidly branching `spectral tree' of possible decimation paths, yielding  $M^{n_{\text{b}}}$ approximate eigenstates once all the $n_{\text{b}}$ bonds have been decimated. The exponentially difficult task of constructing all these possible states can be {partially mitigated by Monte Carlo sampling the spectral tree~\cite{PhysRevX.4.011052}. However, we will show that such costly numerical sampling can be circumvented by working in the limit of infinite effective ``temperature" where all eigenstates are sampled equiprobably -- enabling closed-form analytic results. For entropic reasons, in  large systems this effective $T=\infty$ sampling is dominated by states in the center of the many-body spectrum, and hence is sensitive only to highly excited states.} As the RG proceeds, the characteristic energy gap shrinks, and therefore the remaining spins dominate dynamics on increasingly long time scales. This connects excited-state RSRG (RSRG-X) to a related approach where the decimation is implemented directly on the dynamics by  `integrating out' fast spins~\cite{PhysRevLett.110.067204}.

\vspace{4pt}\noindent{\bf Models - } 
Our goal is to construct fixed-point Hamiltonians for the RSRG-X. To this end, a convenient sequence of models is provided by chains of SU(2)$_k$ anyonic ``spins" - that
transform under ``deformations'' of the SU(2) symmetry of Heisenberg spins. Their $T=0$ properties have previously been analyzed using RSRG techniques, and they provide a convenient language in which the RG rules are transparent and simple to implement.
The simplification is clear for the Ising ($k=2$) case, where working with SU(2)$_2$ anyons (Majorana fermions related to spins by the familiar Jordan-Wigner transformation) places two-spin exchange and on-site fields on equal footing.
As we show below, each value of the integer parameter $k$ that appears in these models labels  a distinct QCG,  with distinct scaling behavior.  The `deformation parameter' $1/k$ measures how far the model is from the $k\rightarrow\infty$ limit of  SU(2) (Heisenberg) spins.

The irreducible representations of SU(2)$_k$ are  labeled by their ``spin" $j=0,\frac{1}{2},1, \dots, \frac{k}{2}$, and obey  
  modified fusion rules truncated at level $k$: $j_1 \otimes j_2 = |j_1-j_2| \oplus \dots \oplus \text{min}(j_1+j_2,k-(j_1+j_2))$. Such SU(2)$_k$ algebras arise naturally in the context of topological quantum computation~\cite{RevModPhys.80.1083}. Consider a one-dimensional chain of  such anyons~\cite{PhysRevLett.98.160409,Trebst01062008} with random couplings~\cite{Bonesteel,FidkowskiPRB08,FidkowskiPRB09}, each with topological charge $S =\frac{1}{2}$ and Hilbert space $\mathscr{H}$ given by the set of fusion outcomes. Owing to the truncated fusion rules,  $\mathscr{H}$ for  $N$ anyons cannot be written as a tensor product of local Hilbert spaces.
   
    \begin{figure}
\includegraphics[width=1\columnwidth]{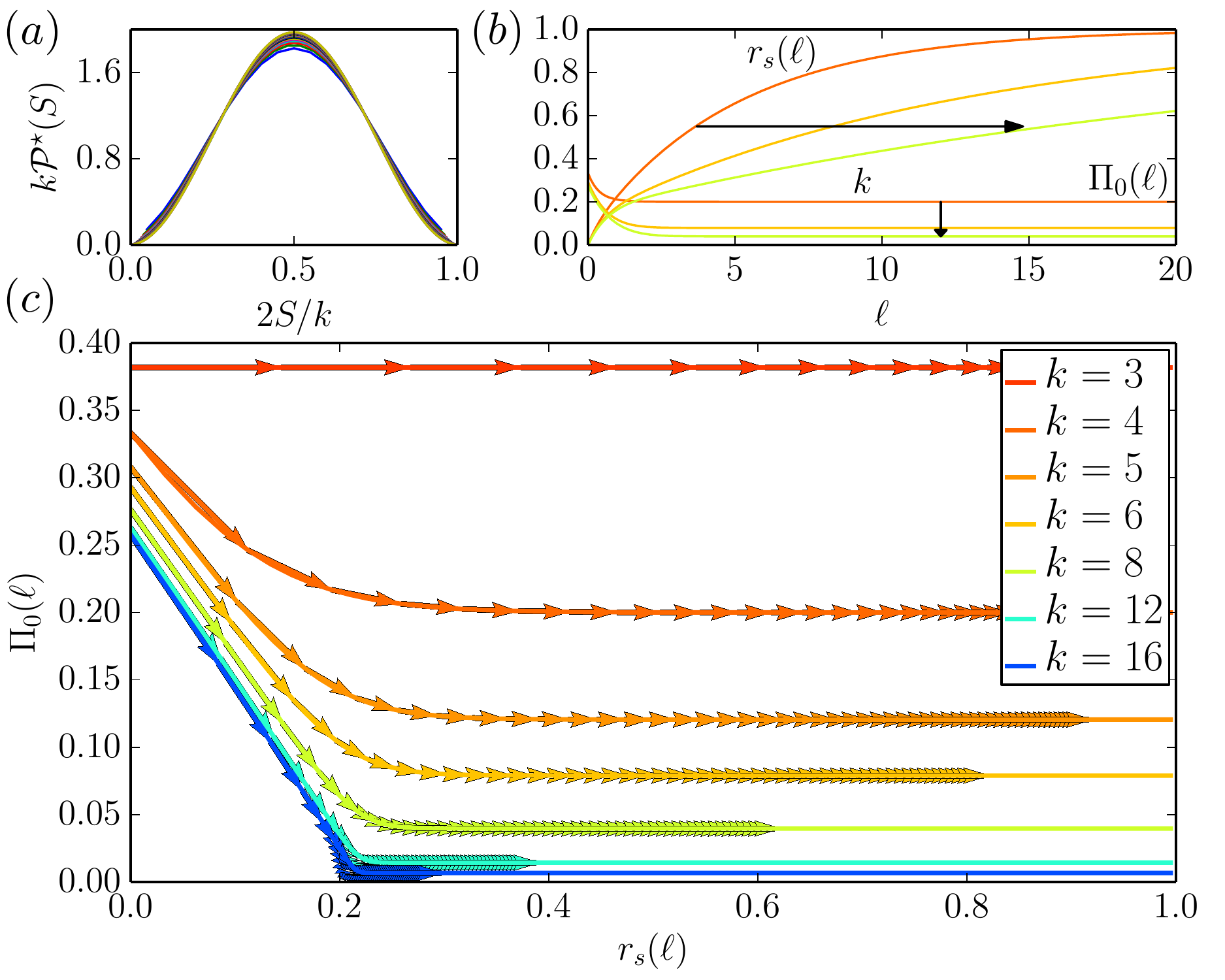}\vspace{-.1in}
\caption{\label{fig:RGflows} {\bf RG Results for Anyon Chains}, obtained by integrating  \eqref{eq:psflow} and \eqref{eq:rsflow} for an initial distribution $\mathcal{P}(S, 0) = \delta_{S,\frac{1}{2}}$. {\bf (a)} Fixed-point spin distribution $\mathcal{P}^*(S)$ as $k\rightarrow\infty$ has universal scaling independent of $\mathcal{P}(S, 0)$. {\bf (b)} {$r_s(\ell)$ (top) $\Pi_0(\ell)$ (bottom) for different $k$ (the arrows indicate increasing values of $k$).}  {\bf (c)} RG trajectories for different $k$, arrow length gives speed of flow. }
\vspace{-.1in}
\end{figure}

Since the renormalization procedure can generate higher effective spins, it is necessary to work with generalizations of this truncated spin-$\frac{1}{2}$ chain that allow the spin $S_i$ on each site to take any value in $\lbrace \frac{1}{2},1, \dots, \frac{k-1}{2} \rbrace$. We work with the Hamiltonian~\cite{FidkowskiPRB09}
\be
H=\sum_i J_i \hat{Q}_i=\sum_i J_i \sum_{S \in S_{i}\otimes S_{i+1}} A_i(S) \hat{P}_S,
\label{eqSUkH}
\ee
where $\hat{P}_S$ is the projector onto the fusion channel $S$ in the fusion $S_{i}\otimes S_{i+1}$. The different fusion channels are weighted by $A_i(S)= \frac{1}{4} (\{S\}_k^2+\{S+1\}^2_k-\{|S_{i+1}-S_i|\}^2_k-\{S_{i+1}+S_i+1\}_k^2)$, where $\{x\}_k = {\sin\left(\frac{\pi x}{k+2}\right)}/{\sin\left(\frac\pi{k+2}\right)}$. 
  Eq.~\eqref{eqSUkH} coincides with Hamiltonian of the SU(2) Heisenberg model  ($H = \sum_{i} J_i \mathbf{S}_1\cdot\mathbf{S}_{2}$) as $k\rightarrow \infty$, since 
 $\mathbf{S}_1\cdot\mathbf{S}_{2} = \frac{1}{2}\sum_{S \in S_{1}\otimes S_{2}} \left[S(S+1) - S_1(S_1+1) - S_{2}(S_{2}+1) \right] \hat{P}_S$.
At strong disorder, each eigenstate of the chain can be constructed by identifying the strongest bond and choosing a fusion channel for its two spins, either forming a singlet or creating a new effective superspin. Crucially, apart from generating different spin sizes, such RSRG-X transformations conserve the form of Hamiltonian~\eqref{eqSUkH}. Since the maximum spin size is bounded by $\frac{k}{2}$, for strong enough initial disorder the RSRG-X scheme flows to strong randomness and becomes asymptotically exact. 

\vspace{4pt} \noindent{\bf RSRG for excited states - }  Starting from~\eqref{eqSUkH}, we can derive the RSRG-X decimation rules explicitly and construct the spectral tree of eigenstates~\cite{Supplement}. At infinite temperature a drastic simplification occurs:  eigenstates are equiprobable, so
  we can construct an average RSRG-X eigenstate by decimating bonds hierarchically and choosing a fusion outcome at each node of the spectral tree with a probability proportional to the number of branches descending from this node, given by the ``quantum dimension" $d_S \equiv \{2S+1\}_k$ (analogous to the $2S+1$ degeneracy of a Heisenberg spin $S$).
The RG flow of the spin distribution decouples from that of the bonds, enabling a simple computation of eigenstate entanglement. This is in marked contrast to the Monte Carlo sampling required at finite temperature,   which entails performing {\it all} decimations (involving {\it both} spin and coupling distributions) leading to a single eigenstate in order to implement the Metropolis algorithm~\cite{PhysRevX.4.011052}.

{Let $\ell$ be the RG depth, defined such that the density of spins at depth $\ell$ is $n(\ell)=e^{-\ell}$}. By examining the distribution of undecimated (super)-spins, ${\cal P}(S,\ell)$,  and change in total number of spins upon fusing two spins either into a superspin or a singlet~\cite{Supplement}, we find that at RG depth $\ell$, 
\be\label{eq:psflow}
\frac{d {\cal P}(S)}{d \ell} = \frac{1}{1+\Pi_0(\ell)}\left[ K_S(\ell) - {\cal P}(S,\ell)(1-\Pi_0(\ell))\right],
\ee
where $K_S(\ell) = \sum_{\substack{S_1,S_2 \neq 0, \frac{k}{2}}} \frac{d_S {\cal P}(S_1){\cal P}(S_2)}{d_{S_1} d_{S_2} }\delta_{S \in S_1 \otimes S_2}$
 is the weighted probability of generating spin $S$, and $\Pi_0 = K_0+K_{k/2}$ is the probability of generating a singlet (note that $k/2$ is also an SU(2)$_k$ singlet).   The spin distribution $ {\cal P}^\star(S)$ at the fixed point is thus given by $K^\star_S = {\cal P}^\star(S)(1-K^\star_0 - K^\star_{k/2})$. Solving this, we find that ${\cal P}^\star(S) = d_S^2/\sum_{S^\prime \neq 0, \frac{k}{2}}  d_{S^\prime}^2 $. For large $k$, the fixed point spin distribution has the  scaling form $ {\cal P}^\star(S) = \frac{1}{k} f(\frac{S}{k/2})$, with $f(x) =2 \sin^2 \pi x$. 
 
We also define the singlet participation ratio $r_s(\ell)$, {\it i.e.} the probability that an original (`UV scale') spin resides in a singlet at depth $\ell$. For 
$r_s(\ell)\sim 1$, all spins of various sizes are essentially in singlets. As $r_s$ depends both on the probability of generating a singlet at depth $\ell$ {\it and} the probability that the spin is not {already} in a singlet~\cite{Supplement}, 
\be\label{eq:rsflow}
\frac{d r_s}{d \ell} = \Pi_0(\ell) (1-r_s(\ell)).
\ee
These flow equations admit straightforward solution for any initial spin distribution (see Fig.~\ref{fig:RGflows}). The spin distribution $\mathcal{P}(S)$ flows very quickly to its fixed-point value $\mathcal{P}^\star(S)$ (indicated by the plateau in $\Pi_0$). Subsequently, $r_s$ grows  towards its fixed-point value. In the plateau, $\Pi_0(\ell) \approx \Pi_0^\star$ where $(\Pi_0^\star)^{-1} \sim \alpha k^3$ for large $k$, with the universal pre-factor $\alpha = (4 \pi^2)^{-1} $. The singlet participation ratio near the fixed point is therefore given by $r_s(\ell) \approx 1 - {\rm e}^{-\Pi_0^\star \ell}$ so that all UV spins are in singlets when $\Pi_0^\star \ell \gg 1$. 

\begin{figure}[ttb]
\begin{center}
\includegraphics[width = 0.8\columnwidth]{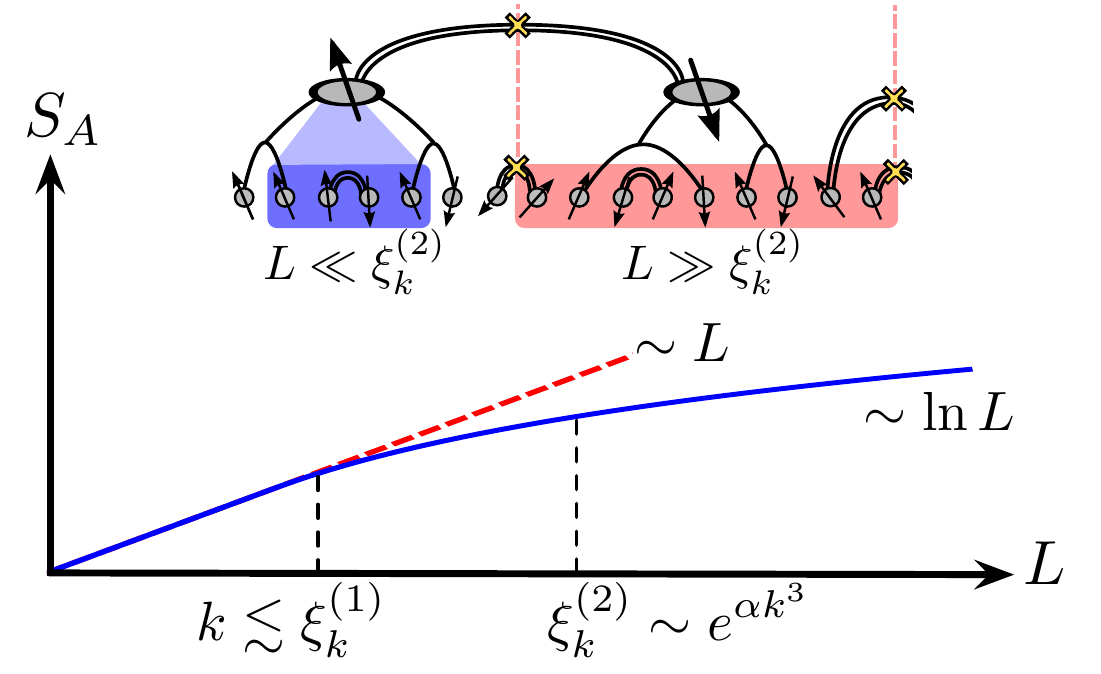}
\vspace{-.3in}
\end{center}
\caption{{\bf Excited state entanglement.} We calculate the entanglement entropy $S_A$ by counting singlets (top). $S_A$ exhibits a crossover from volume-law ($\sim L$) scaling for $L\ll \xi^{(1)}_k$ with $\xi^{(1)}_k \gtrsim k$ to logarithmic scaling for $L\gg\xi^{(2)}_k \sim e^{\alpha k^3}$ with $\alpha=1/4\pi^2$. 
}
\label{fig:entcalc}
\end{figure}

\vspace{4pt} \noindent{\bf Entanglement - } {In order to investigate the dynamics and address the question of thermalization, 
we focus on the spatial dependence of the entanglement entropy  $S_{A}(\ket{\Psi_a})$ in the excited eigenstates $\ket{\Psi_a}$ of (\ref{eqSUkH}). } Specifically, 
we define a measure of the entanglement entropy of a subsystem  $A$ of  size $L$ at infinite effective temperature, $S_{A} =\frac{1}{{\mathscr N}} \sum_{a=1}^{\mathscr N} S_{A}(\ket{\Psi_a})$ with ${\mathscr N} = {\rm dim} \mathscr{H}$, though we expect any suitable average over choice of sub-region, disorder configuration, or eigenstates $\ket{\Psi_a}$ to give similar characterizations.
Eigenstate thermalization requires that $S_{A}$ scale with the volume of $A$, $S_A\sim L$. Sub-volume scaling indicates non-ergodic dynamics implying the system does not self-thermalize in isolation.

The random sampling RSRG-X procedure naturally lends itself to a computation of the $T\rightarrow\infty$ entanglement entropy, averaged over disorder. Consider an interval $A$ of the chain, of length $L$. For $\Pi^\star_0 \ell \gg 1$, or equivalently $L \gg \xi^{(2)}_k \sim {\rm e}^{\alpha k^3}$ with $\alpha = (4 \pi^2)^{-1}$, almost all the original UV spins inside $A$  reside in singlets. 
Region $A$ is entangled with the rest of the system by singlets that cross its boundary. The RG flow yields a constant number of boundary-crossing singlets added at each $\ell$. 
Thus, up to a constant prefactor~\cite{RefaelMoore,FidkowskiPRB08}, $S_A$ is given by counting singlet entanglement up to scale $\ell=\ln L$, when all spins in $A$ have been decimated~\cite{RefaelMoore}: 
\be
S_A \underset{L \gg \xi^{(2)}_k}{\approx} \int^{\ln L}\!\!\!\!\!\!d\ell\!\!\!  \sum_{S\neq 0, \frac{k}{2}}\!\!\frac{{\cal P}(S, \ell)\left[{\cal P}(S, \ell) + {\cal P}\left(\frac{k}{2}-S, \ell\right) \right]}{d_S^2} \ln d_S.\nonumber\\
\label{eqSsinglets}
\ee
The integrand in~\eqref{eqSsinglets} quickly saturates to its constant finite fixed point value, so that the excited-state entanglement scales as $S_A \sim \ln L$ in this regime.

If on the other hand $L$ is much less than $\xi^{(1)}_k$, defined as the scale at which the system starts noticing the finite $k$ truncations ($\xi^{(1)}_k \gtrsim k$), the maximum superspin size grows with $L$, so that one recovers the Heisenberg model in this limit, which we argue below is thermal. 
In other words, the system looks thermal $S_A \sim L$ for $L \ll \xi^{(1)}_k$, but for large intervals $L \gg \xi^{(2)}_k$ (universal regime), the excited-state entanglement entropy scales logarithmically (see Fig.~\ref{fig:entcalc}), characteristic of random-singlet like `critical points' at zero~\cite{RefaelMoore} or finite~\cite{2014arXiv1405.1817H} energy density~\footnote{We do not explicitly rule out a more complicated crossover in the intermediate regime $ \xi^{(1)}_k \ll L \ll \xi^{(2)}_k$.}.

We observe that in the $k\rightarrow\infty$ Heisenberg limit, the crossover scales $\xi^{(1)}_k$, $\xi^{(2)}_k$ diverge, and the RG procedure breaks down  due to the growth of large spins~\cite{Supplement}. Intuitively, we can imagine performing the RSRG until it is on the verge of breaking down. This results in a renormalized chain of large, weakly coupled spins, which behave effectively classically and hence can be expected to thermalize~\cite{Supplement,PhysRevB.80.115104}. This is in accord with other suggestive arguments~\cite{PhysRevLett.110.067204, PhysRevB.89.144201} that  point to ergodicity,  but now we have recovered this from an RG approach that remains controlled at any finite $k$ as the Heisenberg limit is approached, and breaks down precisely at $k=\infty$ when the spins are allowed to grow without bound.

\vspace{4pt}\noindent{\bf Critical Scaling - }
 In order to extract the {dynamical} scaling at the fixed point, we must understand the RSRG flow of the distribution of coupling strengths. This requires going beyond the simplified RG equations for the spin distributions discussed so far. It is convenient to define an energy scale $\Omega$ for the RG, set by the strength of the strongest remaining bond. In units where $\Omega =1$ at the start of the RG,
  the fixed-point probability distribution of bond strength $\beta_i = \log\frac{\Omega}{|J_i|}$ at  scale $\Gamma \equiv \log \frac{1}{\Omega}$ is $\rho(\beta,\Gamma) \equiv \frac{1}{\Pi_0^\star \Gamma} e^{-\beta/\Pi_0^\star\Gamma}$, using standard RSRG techniques~\cite{Supplement}. Intuitively, the scaling is controlled by singlet decimations (and hence $\Pi_0^\star$) as these are the only ones that renormalize the $\rho(\beta,
 \Gamma)$ towards strong disorder. At scale $\Gamma$, the change in the density of remaining spins $n(\Gamma)=e^{-\ell}$ 
depends on the number of spins removed while decimating the strongest bonds (with $J=\Omega$), so that
  \be
  \frac{d\ell}{d\Gamma}= \rho(0,\Gamma) (1 + \Pi_0^\star).
  \ee
Solving this at the fixed point, we find that the typical distance between surviving spins increases as $\sim \(\Gamma/\Gamma_0\)^{1/\psi}$ with $\psi=\Pi_0^\star/(1+\Pi_0^\star)$, implying glassy scaling between time ({$t \sim \Omega^{-1}$}) and distance: {$L^\psi \sim \log t$}.

The tunneling exponent $\psi$ is in general {\it distinct} from the $T=0$ value obtained from the ground state RSRG. For the SU(2)$_k$ models, $\psi^{\text{GS}} = \frac{1}{k}$~\cite{FidkowskiPRB09}, which clearly differs from the results computed from the fixed points in Fig.~\ref{fig:RGflows}. This may be traced to the difference in fixed-point singlet formation probabilities between $T=0$ and $T=\infty$. The sole exception to this is the $k=2$ (Ising) case, where, since all possible decimations lead to singlets by our definition, $\Pi^*_0 = 1$ and hence $\psi=\frac{1}{2}$ {\it independent} of the temperature. Though we defer a detailed analysis to future work~\cite{VPP_Unpub}, we note that in many cases these fixed points apparently have no relevant perturbations and thus represent stable critical {\it phases}, rather than fine-tuned critical {\it points}.

\vspace{4pt} \noindent{\bf Discussion - } We have constructed a set of infinite-randomness fixed points and corresponding scaling limits that control the dynamics of highly excited states of SU(2)$_k$ spin chains at any $k$. These include the fixed points that control the dynamics of disordered Ising and  three-state Potts models that can be mapped to the $k=2$ and $k=4$ chains, respectively~\cite{VPP_Unpub}. We thus uncover an infinite sequence of infinite-randomness critical points/phases that are $T=\infty$ analogs of the Damle-Huse fixed points~\cite{DamleHuse}. In all cases except the Ising model, non-zero energy density is a relevant perturbation that takes the system to a new fixed point with different critical exponents. {While analytic results were obtained for states in the center of the many-body spectrum, corresponding to infinite effective temperature, we expect the universal scaling properties to hold for all eigenstates with arbitrary non-zero energy density, $\e$, due to the following reasoning.}

To target states with energy density $\e$, the RSRG-X can be approximately split into two stages.  First, for RG energy scale $\Omega\gg \e$, essentially all 
decimations yield singlets as in ground state RSRG, resulting in a renormalized chain of predominantly $S=1/2$ spins at scale $\Omega\approx \e$. 
The remaining flow for $\Omega\ll \e$ has all couplings much weaker than temperature and should essentially follow the $T\rightarrow \infty$ behavior described above. {This intuitive argument establishes that the excited state QCG phases persist through almost all states in the many-body spectrum, excepting the ground-state and a set of measure zero low lying excited states. We leave a more detailed study of universal aspects of the crossover for $\Omega\sim \e$, as well as the possibility of energy density tuned delocalization transitions between QCGs and self-thermalizing ergodic phases for future work.}

{\noindent\bf Acknowledgements.}  We are very grateful to D.~Huse for pointing out a significant error in an earlier version of this paper whose correction modified our results while leaving our conclusions unchanged. We thank  J.E. Moore, G. Refael, S.L. Sondhi and A. Vishwanath for insightful discussions and R.~Nandkishore for useful comments on the manuscript. We acknowledge support from the Quantum Materials program of LBNL (RV), the Gordon and Betty Moore Foundation (ACP), and UC Irvine startup funds (SAP). RV thanks UC Irvine for hospitality during completion of this work.

\bibliography{MBL}

\bigskip

\onecolumngrid

\newpage



\section*{Supplementary material}

# Supplemental Material for "Quantum criticality of hot random spin chains"

R. Vasseur,[1,2] A. C. Potter,[1] and S. A. Parameswaran[3]

[1]*Department of Physics, University of California, Berkeley, CA 94720, USA*
[2]*Materials Science Division, Lawrence Berkeley National Laboratories, Berkeley, CA 94720*
[3]*Department of Physics and Astronomy, University of California, Irvine, CA 92697, USA*
(Dated: May 6, 2015)

## I. $SU(2)_k$ ANYONS AND DECIMATION RULES

For the purposes of this paper, $SU(2)_k$ may be viewed as a "quantum deformation" of $SU(2)$ that serves to bound the spin size in a controlled way. The irreducible representations of $SU(2)_k$ are labeled by their "spin" $j = 0, \frac{1}{2}, 1, \ldots, \frac{k}{2}$, and obey modified fusion rules 'truncated at level $k$': $j_1 \otimes j_2 = |j_1 - j_2| \oplus \cdots \oplus \min(j_1 + j_2, k - (j_1 + j_2))$. We study one-dimensional chains of such 'anyons'[2,3] with random couplings[4–6], each with topological charge $S = \frac{1}{2}$. The Hilbert space $\mathscr{H}$ of the chain is then spanned by the basis states $|\ldots j_{i-1} j_i \ldots\rangle$ where $j_{i-1}, j_i \in \{0, \frac{1}{2}, 1, \ldots, \frac{k}{2}\}$ and $j_i$ is contained in the fusion product of $j_{i-1}$ with $S = \frac{1}{2}$. Owing to the truncated fusion rules, $\mathscr{H}$ for $N$ anyons cannot be written as a tensor product of local Hilbert spaces, in contrast to conventional $SU(2)$ spins.

### A. Fixed point Hamiltonians

Since the RG rules can generate arbitrarily high spins, is necessary to work with generalizations of this truncated spin-$\frac{1}{2}$ chain that allow the spin $S_i$ on each site to take any value in $\{\frac{1}{2}, 1, \ldots, \frac{k-1}{2}\}$, and consider configurations in which $j_i \in j_{i-1} \otimes S_i$. We work with the Hamiltonian introduced in Ref. 6,

$$H = \sum_i J_i \hat{Q}_{i,i+1} = \sum_i J_i \sum_{S \in S_i \otimes S_{i+1}} A_i(S) \hat{P}_S, \tag{S1}$$

where $\hat{P}_S$ is the projector onto the fusion channel $S$ in the fusion $S_i \otimes S_{i+1}$, that can be expressed in the basis introduced above using "$F$-moves" (to be defined shortly). The operator $\hat{Q}$ is the same as the truncated dot product introduced in the main text. The different fusion channels are weighted by $A_i(S) = \frac{1}{4}(\{S\}_k^2 + \{S+1\}_k^2 - \{|S_{i+1} - S_i|\}_k^2 - \{S_{i+1} + S_i + 1\}_k^2)$, where $\{x\}_k = \sin\left(\frac{\pi x}{k+2}\right)/\sin\left(\frac{\pi}{k+2}\right)$. (S1) recovers the Heisenberg chain as $k \to \infty$, since $\mathbf{S}_1 \cdot \mathbf{S}_2 = \frac{1}{2} \sum_{S \in S_1 \otimes S_2} [S(S+1) - S_1(S_1+1) - S_2(S_2+1)] \hat{P}_S$. While Eq.(S1) was originally introduced in Ref. 6 as an RSRG-invariant Hamiltonian for anyons, the connection to the Heisenberg model as $k \to \infty$, to the best of our knowledge, has not been previously noted.

It is convenient to introduce a graphical depiction of the Hilbert space in which the fusion rules are simply represented by the merging of particle lines labeled by a particular spin representation. For $SU(2)_k$, in contrast to more complicated theories, there is a single way to fuse an anyon into each of its outcomes. We may represent the 'link basis' states that span the Hilbert space pictorially by a trivalent graph whose 'legs' are the site spins, and whose links are labeled by the possible fusion outcomes of the spins to the left (Fig. S1). The projection operator onto a particular fusion channel of a pair of anyons has a simple graphical representation, also shown in Fig. S1, upto an undetermined constant that is fixed by normalization of the projector. Such fusion diagrams may

be simplified by using a set of simple identities: the 'no tadpole' rule that dictates the vanishing of any graph with a closed loop that can be detached by a single bond; the 'loop rule' that allows us to replace a closed loops of anyon $m$ by a factor equal to the quantum dimension $d_m$; and the '$F$-move' that expresses how to relate the two independent fusion paths in combining three particles into a fourth, that involves an $F$-matrix (also called a $6j$-symbol, see *e.g.* Ref. 6). These rules are depicted graphically in Fig. S2.

The bond strengths $J_i$ are strongly random. The precise distribution of $J$ is unimportant, but for concreteness we take the logarithm of couplings to be distributed according to: $P(\log J^{-1}) = \frac{1}{W} e^{-\frac{\log J^{-1}}{W}}$. The parameter $W$ indicates the standard deviation in the order of magnitude of the couplings, $J_i$, and is a convenient measure for the strength of disorder. We start with all spins-$\frac{1}{2}$, but will soon see that spins of arbitrary size are generated in the RSRG procedure.

For strong disorder ($W \gg 1$), each eigenstate of the chain can be accurately constructed via a real-space renormalization group (RSRG) procedure that begins by identifying the strongest bond with strength $J_S$ and choosing a fusion channel for its two spins (Fig. S3a). For strong disorder, $W \gg 1$, the neighboring bonds on the left ($L$) and right ($R$) are typically much weaker than the strong bond ($J_{L/R} \ll J_S$) and are therefore unable to dynamically alter the chosen fusion channel. If the strong bond spins fuse to a singlet ($S = 0, \frac{k}{2}$), then they drop out of future stages of the RG. Virtual excitations of this frozen singlet then mediate effective coupling $\tilde{J} \approx A(S, S_L, S_R) \frac{J_L J_R}{J_S}$, between the neighboring spins. Here $A$ is a $J$-independent function of spin sizes whose precise form will be unimportant at strong disorder. Another possibility is that the strong bond spins fuse to a "superspin" of size $S \neq 0, \frac{k}{2}$, which continues to participate in later stages of the RG and interacts with its neighboring spins via renormalized coupling $\tilde{J}_{L/R} \approx B(S, S_{L/R}) J_{L/R}$. The coefficient $B$ can take either sign, but this will turn out to be unimportant for dynamics in highly excited states. Crucially, apart from generating different spin-sizes, each RSRG step preserves the form of the Hamiltonian (S1).

### B. Decimation rules

We emphasize that the general form of the decimation rules is entirely dictated by the structure of perturbation theory, and that the precise coefficients appearing in the renormalized couplings are not important at large disorder. Nevertheless, we give the explicit form of the decimation rules below for concreteness. This first part of our discussion very closely parallels that of[6], but we reproduce the details here for completeness; we caution the reader that we use a different notation, in particular differing on the definition of $\lfloor \ldots \rfloor_k$ and $\{\ldots\}_k$.

In deriving these decimation rules, we consider the case $k$ odd, where we can restrict ourselves to integer spins to make the calculation a bit simpler, but the final results hold for even $k$ and half-integer spins as well. The easiest way to phrase the decimation rules is in terms of projection operators; in essence, we project the spins on a strong bond into a specific fusion channel and

FIG. S1: **Graphical Notation for Anyons.** (Left) SU(2)$_k$ link basis state; each link may be labeled by a possible fusion outcome of the anyons located on the 'legs' of the figure. (Right) Projector onto the spin-$S$ fusion channel, $\hat{P}_S$; the corresponding normalization $c_S$ is fixed by requiring $\hat{P}_S^2 = \hat{P}_S$.

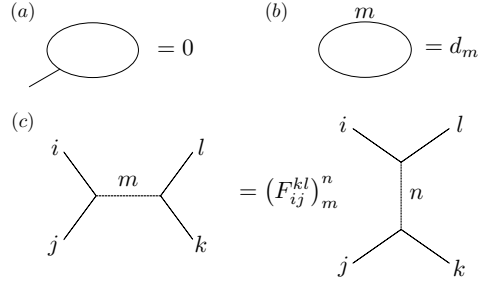

FIG. S2: **Rules for simplifying fusion diagrams.** (a.) 'No tadpole' rule. (b.) 'Loop rule'. (c.) '$F$-move'.

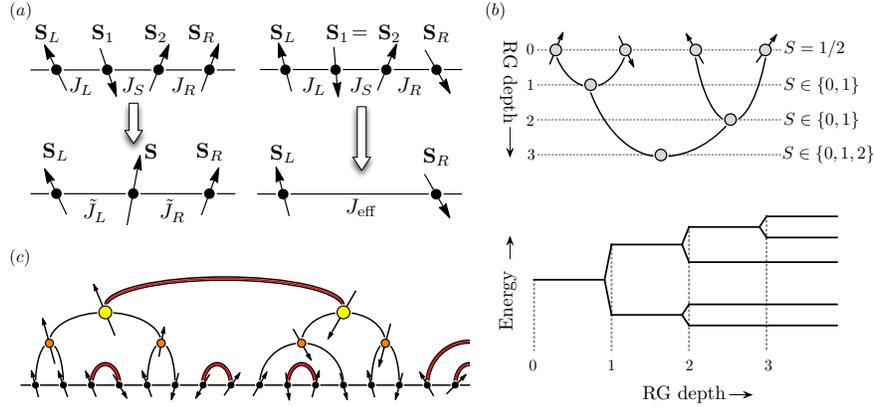

FIG. S3: **RSRG approach to quantum critical glasses.** (a) RG rules for combining two spins $S_1, S_2$ coupled by a strong bond $J_S$ into a new spin $S$, or into a singlet, and corresponding renormalized couplings for adjacent spins $S_L, S_R$. (b) Sketch of RG procedure on a sample 4-site chain demonstrating how at each step there are additional possibilities for the effective spin and the spectral tree obtained from the RG for the same chain. (c) Sketch of typical random-singlet excited state in the quantum critical glass; singlet bonds (thick red lines) are between 'superspins' (colored circles) formed from combining lattice scale spins (black circles) through non-singlet fusion channels (thin black lines).

then compute the coupling of the fused spins with the remaining ones using perturbation theory. This procedure is most conveniently implemented using the graphical representation of the fusion procedure introduced above, and is especially transparent once one realizes that (S1) may be rewritten by inverting an $F$-move, so that the operator $\hat{Q}_{i,i+1}$ represents the exchange of the '1' anyon between sites $i, i+1$ (indeed, this is how the form of $H$ was originally inferred[6].) The graphical depiction of the operator $\hat{Q}_{i,i+1}$ ( rewritten as an anyon exchange) is given in Fig. S4.

We first explain how to handle first-order decimations. Consider the case where two spins $S_1$ and $S_2$ fuse to a new effective spin $S$; we wish to compute, for instance the effective coupling $\tilde{J}_R$ between the spin $S_R$ on the right of $S_2$ and $S$ in terms of $J_R$, the old coupling between $S_R$ and $S_2$ (the results for $\tilde{J}_L$ in terms of $J_L$ are similar and so we do not show them explicitly.) The fusion of $S_1$ and $S_2$ is captured by the projection operator $\hat{P}_S^{12}$, and the corresponding renormalized coupling is given by $\hat{P}_S^{12}\hat{Q}_{23}\hat{P}_S^{12}$ (see Fig. S5(a)). By performing a (rather tedious) sequence of $F$-moves, we find that the coupling between the composite spin $S$ and $S_R$ takes the form $\tilde{J}_R \hat{Q}_{SS_R}$, with

$$\hat{Q}_{i,i+1} = \begin{array}{c} \phantom{i} \phantom{i+1} \\ \big| \quad \underline{\phantom{1}}\!1\!\underline{\phantom{1}} \quad \big| \\ \big| \phantom{xxxxx} \big| \\ i \quad\quad i+1 \end{array}$$

FIG. S4: **The $\hat{Q}$-operator that defines the truncated dot product.** The operator acts on nearest-neighbor sites and can be rewritten as a sum of projectors using an $F$-move. Note that it is nonzero only when the fusion rules are obeyed at its vertices.

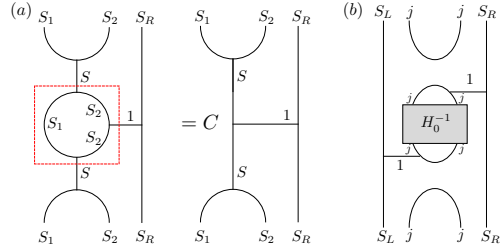

FIG. S5: **Decimation rules in the graphical representation[6].** (a.) First-order decimation when spins $S_1, S_2$ fuse to a new spin $S$. In order to determine the effective Hamiltonian between the composite spin $S$ and $S_R$, we need to compute $\hat{P}_S^{12}\hat{Q}_{23}\hat{P}_S^{12}$; using a a sequence of $F$-moves within the red box, we can show that this is $\tilde{J}_R\hat{Q}_{SS_R}$, i.e. has the same form as the original Hamiltonian. The explicit form of $\tilde{J}_R$ is given in (S2). (b.) Second-order decimation, where spins $S_1 = S_2 = S$ fuse to a singlet. Once again, a series of $F$-moves may be employed and after a tedious computation we can show that the effective Hamiltonian between $S_L, S_R$ is $J_{\text{eff}}\hat{Q}_{S_LS_R}$, with $J_{\text{eff}}$ given by (S6).

$$\tilde{J}_R = \frac{\{S_2+S-S_1\}_k\{S_1+S_2+S+2\}_k - \{S_1+S_2-S\}_k\{S_1+S-S_2\}_k}{\{2S\}_k\{2S+2\}_k}J_R. \tag{S2}$$

In the $k \to \infty$ limit, we recover

$$\tilde{J}_R = \frac{S(S+1) + S_2(S_2+1) - S_1(S_1+1)}{2S(S+1)}J_R, \tag{S3}$$

which concurs with the result for conventional SU(2) Heisenberg spins, as expected.

Let us now imagine that the two spins $S_1 = S_2 = j$ fuse to a singlet with topological charge (spin) 0. If we denote the adjacent spins on either side of the strongly coupled bond to be $S_L, S_R$, then we may write the 4-spin Hamiltonian $H_4$ in suggestive form: $H_4 = H_0 + H'$, where the 'bare' Hamiltonian has a strong bond coupling $S_1, S_2$, and $H'$ denotes the weak residual couplings,

$$H_0 = J_2(\hat{Q}_{23} - \langle\hat{Q}_{23}\rangle), \quad H' = J_L\hat{Q}_{12} + J_R\hat{Q}_{43}, \tag{S4}$$

where $\langle\hat{Q}_{23}\rangle$ denotes the expectation value of $\hat{Q}_{23}$ in the state when the two spins fuse to the trivial state. Using second-order perturbation theory we find that the effective Hamiltonian between $S_L$, $S_R$ takes the form

$$H_{\text{eff}} = -\frac{1}{\Omega}\hat{P}_0^{12}H'(1-\hat{P}_0^{12})H'\hat{P}_0^{12}, \tag{S5}$$

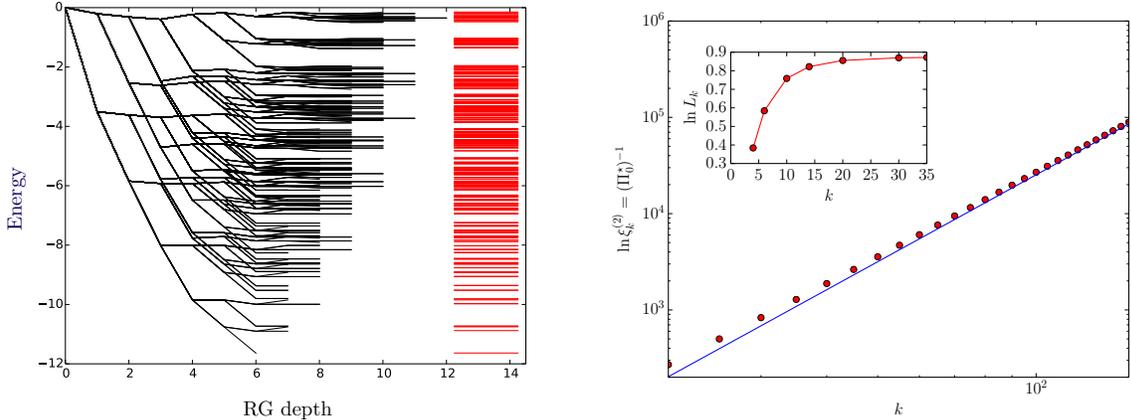

FIG. S6: Left: Example of RG tree construction of the many-body eigenstates of a strongly disordered Fibonacci anyonic chain ($k = 3$), and comparison to exact diagonalization results (in red). Right: $k$ dependence of the crossover scale $\xi_k^{(2)}$, the blue line is the prediction $\xi_k^{(2)} \approx e^{k^3/(4\pi^2)}$. Inset: Length scale $L_k$ at which $\Pi_0(\ell)$ saturates to its fixed point value $\Pi_0^\star$ as a function of $k$: $\ell_k = \ln L_k$ grows very slowly with $k$.

as depicted graphically in Fig. S5(b). After a tedious calculation, we find that once again we may write the Hamiltonian between the residual spins in the original form, $H_{\text{eff}} = J_{\text{eff}} \hat{Q}_{S_L S_R}$, with

$$J_{\text{eff}} = \frac{2}{\{3\}_k} \frac{J_L J_R}{\Omega} \tag{S6}$$
$$\times \frac{\{2j\}_k \{2j+2\}_k}{\{2j\}_k \{2j+2\}_k - \{2j-1\}_k \{2j+3\}_k + 1}.$$

In the Heisenberg limit, this becomes

$$J_{\text{eff}} = \frac{2}{3} j(j+1) \frac{J_L J_R}{\Omega}. \tag{S7}$$

Whereas the $j$-dependent prefactor is bounded by $\sim k^2$, the factor of $\frac{J_L J_R}{\Omega}$ has typical value $e^{-(1+\Pi_0^\star)\Gamma}$ at RG energy scale $\Gamma$. Hence for $\Gamma \gg \log k$, one can neglect the $j$-dependent prefactor in the renormalized couplings. Thus, for finite $k$ there is no distinction in the asymptotic RSRG-X flows between decimating the bond with the strongest coupling, $J_i$, or that with the largest energy gap, $\sim j(j+1)J_i$. This allows us to separate the flow of spin- and bond- distributions below.

## II. FLOW EQUATIONS AND STRONG DISORDER FIXED POINT

Iterating the renormalization procedure generates a tree of fusion choices (Figs. S3b and S6a), each branch of which corresponds to a different many-body eigenstate. Sampling these eigenstates equiprobably results in enormous simplifications that enable analytic progress. This averaging is analogous to working at infinite temperature, and for large systems is overwhelmingly dominated by states in the middle of the many-body spectrum (as there are exponentially more states there than at any other energy for entropic reasons, c.f. Fig. S7). Such equiprobable sampling of eigenstates (tree branches) is accomplished by weighting each fusion choice (tree node) by the fraction of branches descending from this node. Specifically, two spins $S_1, S_2$ are fused to spin $S$ with probability proportional to the quantum dimension $d_S$ of the new spin.

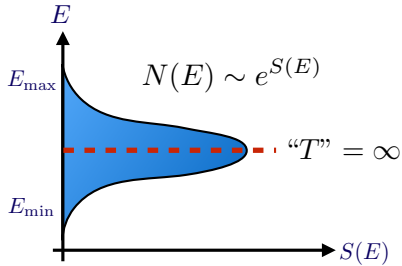

FIG. S7: **Infinite "Temperature" Sampling.** The microcanonical entropy, $S$, as a function of many-body energy, $E$, is expected to be sharply peaked (e.g. adopting a Gaussian profile) at the center of the many-body spectrum. Since the number of many body states scales as $N(E) \sim e^{S(E)}$, almost all states are contained near this entropy peak in the limit of large system size, and hence our equiprobable sampling RSRG-X scheme is dominated by states near this energy.

### A. Spin distribution

This RSRG procedure generates superspins of various sizes, characterized by a distribution $\mathcal{P}(S)$ that will depend on the RG scale $\ell$ introduced in the main text. To derive the flow equation for the spin distribution, we start with $N$ spins of various sizes and perform $n$ RG steps. Each step follows a fusion channel with a probability $p_{S_1 S_2}^S = \frac{d_S}{d_{S_1} d_{S_2}} \delta_{S \in S_1 \otimes S_2}$, where $d_S = \{2S+1\}_k$ is the quantum dimension of anyon $S$. The change in the number of spins of size $S$ is then $\Delta N(S) \approx n \left[ -2\mathcal{P}(S) + \sum_{S_1, S_2} p_{S_1 S_2}^S \mathcal{P}(S_1) \mathcal{P}(S_2) \right]$, where $\mathcal{P}(S) = N(S)/N$ is the fraction of spins of size $S$ before implementing the RG steps. This implies that the probability distribution changes by

$$\frac{\Delta \mathcal{P}(S)}{n/N} = \sum_{S_1, S_2} p_{S_1 S_2}^S \mathcal{P}(S_1) \mathcal{P}(S_2) - \mathcal{P}(S) \left[ 1 - \sum_{S' \neq 0, \frac{k}{2}} \frac{\mathcal{P}(S')^2}{d_{S'}^2} - \sum_{S' \neq 0, \frac{k}{2}} \frac{\mathcal{P}(S') \mathcal{P}\left(\frac{k}{2} - S'\right)}{d_{S'} d_{\frac{k}{2} - S'}} \right]. \quad (S8)$$

This gives the change in the probability distribution with respect to the number of RG steps. We would like to rewrite this expression in terms of the log of the number of spins left compared to the number of spins we started with: $\ell = -\ln N'/N_0$. Proceeding analogously to the previous calculation, we find that $\Delta \ell = -\ln \frac{N + \Delta N}{N} \approx -\frac{\Delta N}{N}$ can be expressed as

$$\Delta \ell = \frac{n}{N} \left[ 1 + \sum_{S \neq 0} \frac{\mathcal{P}(S)^2}{d_S^2} + \sum_{S' \neq 0, \frac{k}{2}} \frac{\mathcal{P}(S') \mathcal{P}\left(\frac{k}{2} - S'\right)}{d_{S'} d_{\frac{k}{2} - S'}} \right]. \quad (S9)$$

Assembling these expressions, we find

$$\frac{d\mathcal{P}(S)}{d\ell} = \lim_{n \to 0} \frac{\Delta \mathcal{P}(S)}{\Delta \ell} = \frac{1}{1 + \Pi_0(\ell)} \left[ K_S(\ell) - \mathcal{P}(S) \left(1 - \Pi_0(\ell)\right) \right], \quad (S10)$$

as claimed in the main text.

### B. Singlet Participation Ratio

We now compute the flow of the singlet participation ratio $r_s(\ell)$, defined as the probability that a microscopic (UV scale) spin resides in a singlet at RG depth $\ell$. When this becomes $\mathcal{O}(1)$, we are

in the singlet dominated regime. To this end, we define the following quantities. Let $N_0$ be the initial number of spins and $N(\ell)$ be the remaining number of super-spins after running the RG for length $\ell$, and let $\bar{s}(\ell)$ be the average number of microscopic spins that make up a surviving "super-spin" at scale $\ell$. After change in RG scale $\Delta\ell$, the change in the number of spins participating in singlets is given by

$$\Delta r_s(\ell) N_0 = \underbrace{\frac{1}{2}}_{\text{2 spins/singlet}} \times \underbrace{\left[N_0 e^{-\ell} \Delta \ell\right]}_{\text{\# spins decimated}} \times \Pi_0(\ell) \times \underbrace{(2\bar{s}(\ell))}_{\text{\# UV spins involved in new singlet}}, \quad (S11)$$

from which we find that

$$\frac{dr_s}{d\ell} = e^{-\ell} \Pi_0(\ell) \bar{s}(\ell). \quad (S12)$$

On the other hand, at depth $\ell$, there are $N_0 e^{-\ell}$ spins remaining, which are formed from $(1 - r_s(\ell)) N_0$ UV spins. Therefore, the typical number of microscopic spins making up each surviving super-spin is:

$$\bar{s}(\ell) = (1 - r_s(\ell)) e^{\ell}, \quad (S13)$$

Combining the above gives

$$\frac{dr_s}{d\ell} = \Pi_0(\ell) (1 - r_s(\ell)). \quad (S14)$$

### C. Fixed-point spin distribution

The spin distribution, $\mathcal{P}(S, \ell)$, quickly asymptotes to a steady-state form: $\mathcal{P}^\star(S) = d_S^2 / \sum_{S' \neq 0, k/2} d_{S'}^2$ as discussed in the main text. We find that one can deduce most of the important properties of the strong disorder phase from the spin distribution $\mathcal{P}^\star(S)$ alone. From $\mathcal{P}^\star(S)$ it is straightforward to compute the probability of decimating a strong bond into a singlet (fusing to spin 0 or $\frac{k}{2}$), which asymptotes to $\Pi_0^\star = \left(\frac{1}{2} \sum_{S \neq 0, k/2} d_S^2\right)^{-1}$. This quantity plays a key role in determining scaling properties of the strong disorder phase. In particular, it dictates the $k$-dependence of the crossover scale $\xi_k^{(2)}$ defined as the scale at which the singlet participation ratio $r_s(\ell)$ becomes $\mathcal{O}(1)$. Since $r_s(\ell) \approx 1 - e^{-\ell \Pi_0^\star}$ near the fixed point, we have $\ln \xi_k^{(2)} = (\Pi_0^\star)^{-1}$. Using the universal scaling of the spin distribution $\mathcal{P}^\star(S) = \frac{1}{k} f(2S/k)$ at the fixed point, with $f(x) = 2 \sin^2 \pi x$, we find that $\Pi_0^\star$ for large $k$ scales as $(\Pi_0^\star)^{-1} \underset{k \to \infty}{\sim} \frac{k^3}{2\pi^2} \int_0^1 dx \sin^2(\pi x)$. Therefore,

$$\xi_k^{(2)} \sim \exp\left(\alpha k^3\right), \text{ with } \alpha = \frac{1}{4\pi^2} \approx 0.0255. \quad (S15)$$

The universal prefactor $\alpha$ is completely characterized by the spin distribution $\mathcal{P}^\star$ at the fixed point. The resulting eigenstate for $L \gg \xi_k^{(2)}$ has a strongly random pattern of singlets formed between super-spins, with a broad distribution of singlet bond sizes (see Fig. S3c). This random-singlet state has all the properties of a zero-temperature random quantum critical point: for example, it exhibits power law scaling of (disorder averaged) correlation functions[7,8], and critical scaling of (disorder averaged) entanglement entropy for sub-regions of size $L$: $S(L) \sim \log L$.[9] Remarkably, we find that *every eigenstate* in the spectrum of the anyon chains exhibits these quantum critical properties typically associated only with zero-temperature quantum phases. Hence, we see that the strong disorder phase of these random anyonic spin chains does not reach thermal equilibrium,

and is not governed by the laws of thermodynamics. We refer to such a non-ergodic system as a "quantum critical glass".

Note also that $\Pi_0(\ell)$ saturates to its fixed point value $\Pi_0^\star$ with a length scale $L_k = e^{\ell_k}$ that is much smaller than $\xi_k^{(2)}$ (so that the expression $r_s(\ell) \approx 1 - e^{-\ell \Pi_0^\star}$ becomes very quickly justified). Both length scales $\xi_k^{(2)}$ and $L_k$ are plotted in the right panel of Fig. S6.

### D. Coupling distribution and dynamics

A more detailed understanding of the strong disorder phase can be obtained by analyzing how the distribution of bond strengths evolves along the RSRG flow. For this purpose, it is convenient to define an energy scale for the RG, $\Omega$ (we choose units in which $\Omega \equiv 1$ at the start of the RG), and to work with logarithmic variables $\Gamma = \log \frac{1}{\Omega}$ and $\beta_i = \log \frac{\Omega}{|J_i|}$. Denoting the probability of having bond strength $\beta$ at energy scale $\Gamma$ by $\rho(\beta, \Gamma)$ one can derive the evolution of $\rho$ with energy scale $\Gamma$:

$$\frac{\partial \rho(\beta, \Gamma)}{\partial \Gamma} = \frac{\partial \rho(\beta, \Gamma)}{\partial \beta} + \rho(0, \Gamma) \left[ \Pi_0^\star \int_0^\beta d\beta' \rho(\beta', \Gamma) \rho(\beta - \beta', \Gamma) + (1 - \Pi_0^\star) \rho(\beta, \Gamma) \right]. \tag{S16}$$

The first term simply maintains the normalization of the bond probability distribution. The second, bracketed term in (S16) represents the probability that the strongly-bonded spins fuse to a singlet ($\sim \Pi_0^\star$) or residual super-spin ($\sim (1 - \Pi_0^\star)$), thereby producing a renormalized bond-strength $\tilde{J} = e^{-\tilde{\beta}/\Gamma}$. In (S16), we have neglected the spin-dependent prefactors $A, B$, as these are unimportant in the limit of strong disorder. Note that we have also assumed that the spin distribution is already converged to its fixed point value $\mathcal{P}^\star(S)$ so that $\Pi_0^\star$ does not depend on $\Gamma$. (The rapid convergence of $\mathcal{P}$ to $\mathcal{P}^\star$ is clearly observed in the RG flows shown in the main text.) Despite its complicated integro-differential character, (S16) is readily solved by the simple function

$$\rho(\beta, \Gamma) = \frac{1}{\Pi_0^\star \Gamma} e^{-\beta/(\Pi_0^\star \Gamma)}, \tag{S17}$$

similarly to the $T = 0$ case[7,8]. As the flow progresses, the effective disorder strength increases as $W(\Gamma) = \Pi_0^\star \Gamma$ due to the bond renormalizations. Hence we see that as long as the initial disorder is sufficiently strong, the system flows to an infinite-randomness fixed point where the RSRG procedure becomes asymptotically exact. Note that we chose to only keep track of the absolute value of the couplings $J_i$. It is also possible to follow the distributions of both ferromagnetic and antiferromagnetic couplings independently to show that at the fixed point, couplings are either ferromagnetic or antiferromagnetic with equal probability.

The quantities $\mathcal{P}^*(S)$ and $\rho(\beta, \Gamma)$ completely characterize the long-distance universal properties of the strong disorder phase. In particular, the dynamical scaling properties of the strong disorder fixed point are readily extracted from $\rho(\beta, \Gamma)$ and from the relation between the RG scales $\ell$ and $\Gamma$: $d\ell = (1 + \Pi_0^\star) \rho(0, \Gamma) d\Gamma$. As discussed in the main text, this implies the following scaling relation between energy and distance $\Gamma \sim L^\psi$ with $\psi = \Pi_0^\star/(1 + \Pi_0^\star)$. From this, we deduce that the typical coupling between bonds separated by length $L \gg 1$ is

$$J(L) = e^{-\left[\Gamma(L) + \int d\beta \, \beta \rho(\beta, \Gamma(L))\right]} \approx e^{-(1 + \Pi_0^\star)\Gamma_0 L^\psi}, \tag{S18}$$

where $\Gamma_0$ in the UV RG scale corresponding to $L = a$ (lattice spacing). Hence the critical properties of the strong-disorder phase are characterized by the logarithm of energy scaling like a power law in distance, with characteristic exponent $\psi$. This power law scaling of *log-energy* is very different from

the usual power law relation between *energy* and distance found in clean or weakly random critical points, but is typical for infinite-randomness fixed points[5–8,11]. The exponent $\psi$ characterizes the dynamics of the system, and appears for example in the slow entanglement growth $\sim \log^{1/\psi} t$ after a global quench[10,12] – except for the Ising case $k = 2$, see below. Note that the stretched-exponential form of the interactions (S18) can also be used to argue that resonances that could proliferate and lead to delocalization are irrelevant at the infinite disorder fixed point[10].

We also note that $\psi$ tends to 0 in the Heisenberg limit $k \to \infty$. This indicates that the log-time to tunnel through a length $L$ chain of Heisenberg ($k = \infty$) chain is faster than any power-law in $L$. This strongly suggests algebraic dynamical scaling relation between length and time with finite dynamical exponent $z$, and provides further evidence that the Heisenberg chain is delocalized (and hence thermal) for arbitrarily strong disorder, supporting the semiclassical arguments and numerical breakdown of RSRG-X described below.

While our RSRG procedure simplifies dramatically for highly excited states near the center of the many-body spectrum, results for lower-energy eigenstates in the tails of the spectrum with energy density $E$ may be approximately obtained by following the ground-state RSRG procedure to scale $\Omega \approx E$, and then continuing with the highly-excited state RSRG procedure for lower energy scales. This argument suggests that energy density is a relevant perturbation to the ground-state fixed point, and that the dynamical scaling properties exhibit a crossover, with the infinite-randomness exponent evolving from the ground state value $\psi^{\text{GS}} = \frac{1}{2}$ (antiferromagnetic couplings) or $\psi^{\text{GS}} = \frac{1}{k}$ (ferromagnetic couplings) to the excited state result $\psi = \Pi_0^\star/(1 + \Pi_0^\star)$ on characteristic length scales $L \sim \left(\log \frac{1}{E}\right)^{\psi^{\text{GS}}}$. At face value, a more quantitative analysis of the intermediate energy spectrum requires numerical Monte Carlo sampling of the RSRG spectral tree (though analytic simplifications may be possible) and is left for future work.

### E. Special Case - Ising Model

The Ising model for $k = 2$ is exceptional in a two ways. First, the ground-state log-dynamical exponent $\psi_{\text{GS}}$ coincides with the excited state one, $\psi = \frac{1}{2}$. Second, the Ising model is the only member of the family of SU(2)$_k$ which can be mapped to non-interacting fermions. Specifically, we note that, for $k = 2$, as written (S1) is a free-fermion Hamiltonian with a fully integrable spectrum. Strictly in this non-interacting limit, we do not expect a delocalized phase for any disorder strength. To move away from this fine-tuned point, we imagine adding weak interactions (e.g. a $J'\sigma_i^x \sigma_{i+1}^x$ term) that break integrability and produce a non-integrable thermal liquid phase at high temperature and weak disorder. Such perturbations are irrelevant at strong disorder, and do not disturb most of the scaling properties of the quantum critical glass. Interactions are, however, crucial for the dynamics of dephasing, which does not occur in non-interacting localized systems. For the Ising model, therefore, the dephasing dynamics are set by the scaling dimension of the leading irrelevant interaction term[12] rather than the log-dynamical exponent $\psi$ that governs the decay of typical correlation functions. Interactions are also required to obtain an ergodic, thermal phase at weak disorder.

### F. Some remarks on standard numerical methods

We now briefly comment on potential numerical checks of our results. We first point out that numerical methods like the density matrix renormalization group (DMRG) or quantum Monte Carlo are not ideal to study many-body localization properties since they do not give access directly to excited states. Moreover, we expect that Monte Carlo simulations would easily get stuck in

metastable configurations because of the inherent "glassy" nature of the problem. Therefore, exact diagonalization will likely be the most appropriate tool to look for infinite randomness critical properties in highly excited states. Another possibility to verify our results would be to investigate the dynamics of anyonic chains using the time-dependent DMRG (or some other time-evolution method based on matrix product states). In both cases, we emphasize that accessing the universal infinite randomness (localized) regime $L \gg \xi_k^{(2)}$ may be challenging because of the length scale $\xi_k^{(2)}$ that we predict to be of order $\sim 16$ sites already for $k = 3$. It would be very interesting to check whether the excited-state criticality predicted in our work can be observed in the random Fibonacci chain $k = 3$ using these numerical methods.

## III. RSRG-X FOR THE RANDOM-BOND HEISENBERG MODEL

We briefly discuss some of the pitfalls in applying RSRG-X to the random-bond Heisenberg model, As in the anyon case, we allow for the formation of higher spins by considering the on-site spin as an additional random variable, and tracking the RG flow of the spin distribution, in addition to that of the bond strengths (see also[13,14]). The most general decimation step now involves a strong bond between two arbitrary spins $S_1$, $S_2$ that fuse to yield a new spin $S \in \{|S_1 - S_2|, \ldots, S_1 + S_2\}$. Consider the Heisenberg Hamiltonian for four successive spins $\mathbf{S}_L, \mathbf{S}_1, \mathbf{S}_2, \mathbf{S}_R$, given by $H = J_L \mathbf{S}_L \cdot \mathbf{S}_1 + J_S \mathbf{S}_1 \cdot \mathbf{S}_2 + J_R \mathbf{S}_2 \cdot \mathbf{S}_R$, where we assume $J_S \gg J_L, J_R$. Except in the case when two identical spins fuse to a singlet, the effective coupling is obtained by first-order perturbation theory and is $H_{\text{eff}} = \tilde{J}_L \mathbf{S}_L \cdot \mathbf{S} + \tilde{J}_R \mathbf{S} \cdot \mathbf{S}_R$, with $\tilde{J}_{L,R} = \frac{S(S+1) \pm S_1(S_1+1) \mp S_2(S_2+1)}{2S(S+1)} J_{L/R}$. In the case when $S_1 = S_2$ and they form a singlet (Fig. S3(a)), a second-order calculation gives $H_{\text{eff}} = J_{\text{eff}} \mathbf{S}_L \cdot \mathbf{S}_R$, with $J_{\text{eff}} = \frac{2S_1(S_1+1)}{3} \frac{J_L J_R}{J_S}$.

That the resulting RSRG-X procedure is fatally flawed is especially clear at infinite temperature, where we can sample the tree at random, picking each of the fusion outcomes $S$ at each decimation step with a probability proportional to its degeneracy, i.e., $2S + 1$. The resulting flow is straightforward to determine numerically, but we sketch the intuition first. Under the random decimation, we will certainly generate some fraction of large spins. Qualitatively, the formation of increasingly large spins apparently leads to a breakdown of the RG, since a large spin leads to a large effective coupling, contrary to the perturbative justification given for the decimation. However, it is possible that the flow is such that the combination of the spin and the bare couplings that give the effective coupling remains small.

To clarify the situation, we simulated the infinite-temperature RG procedure on a $10^3$-site chain, holding the length fixed by replacing decimated spin(s). The extra spin(s) are added at a distance $L/2$ from the strong bond to avoid any accidental correlations between them and the bond being decimated. The new spins and their couplings are chosen randomly from the spins in the previous distribution, excluding those on or adjacent to the strong bond being fused. For large systems, one expects that these are representative of the true distributions so that this process approximates sampling the latter, but without the additional cost of fitting the spin-distribution at every step of the RG procedure. To preserve the correlation between spins and couplings, when we add a new spin with spin-size $S_i$ equal to that of site $i$ in the chain, we couple the new spin to the rest of the chain with strength equal to $J_{i,i+1}$.

Fig. S8 shows the evolution of mean spin size, $\bar{S}$, and error in neglecting higher-order contributions to $J_{\text{eff}}, \tilde{J}_{L/R}$ over the course of $10^4$ RG steps. Here we define the error for a given RG step as $\frac{J_{\text{eff}}}{J_S}$ or $\frac{(J_{L/R} S_{L/R} S_{1/2})^2}{J_S S^2 \tilde{J}_{L/R}}$ when two spins are fused to a singlet or non-vanishing spin respectively. The data clearly show that the unbounded growth of spin size produces a breakdown in the validity of the RSRG-X scheme. It is natural to expect a resulting delocalized thermal phase, since the RG

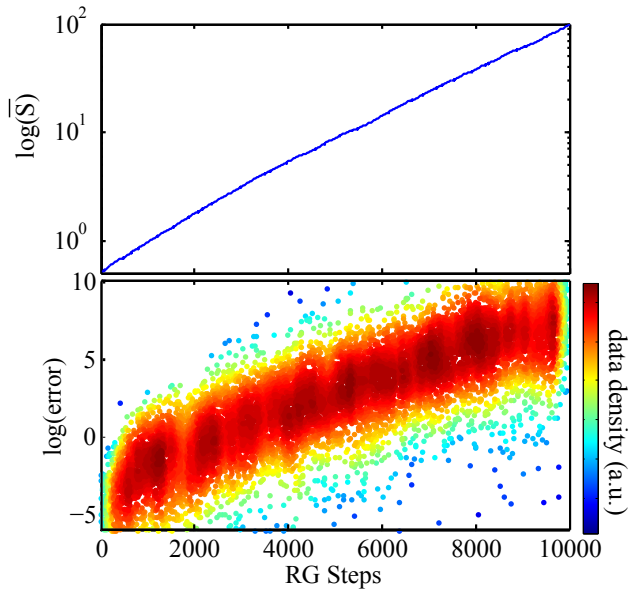

FIG. S8: **Breakdown of RSRG for the Heisenberg model.** The mean residual spin $\bar{S}$ (top) and the RG error (bottom) grow under the RG, indicating its breakdown.

breakdown is caused by a proliferation of resonant bonds (with $J_{L,R} \sim J_S$) that can coherently transport energy.

To gain further insight into the fate of the high-temperature Heisenberg chain, consider running the RSRG-X procedure until it just before it begins to break down. This results in an effective Heisenberg model with a broad distribution of large superspins with weak residual couplings. These large spins behave essentially classically, since the quantum mechanical uncertainty in spin orientation for a spin-$S$ object is suppressed as $1/S$. Conventional wisdom holds that one dimensional spin systems with local interactions (particularly those with isotropic continuous spins) cannot form a glassy state with non-ergodic dynamics[15], further supporting the hypothesis that the random-bond Heisenberg chain thermalizes at high energy density (see also[10] and[16] for similar hints in this direction).

From the preceding discussion, it is clear that the unbounded growth of spins under RSRG-X leads inevitably to the breakdown of perturbation theory. It is tempting to control this growth by simply imposing a cutoff so that spins above a certain $S_{\max}$ are discarded. However, it is straightforward to see that this leads to inconsistencies in the 'fusion rules' for angular momentum addition. Instead, we may truncate $SU(2)$ in a controlled way by "deforming" the group to its quantum version, $SU(2)_k$: in other words, the anyon models studied in the main text serve as truncations of the Heisenberg model. The conclusions that may be drawn from taking their $k \to \infty$ limit are discussed in the main text.

---



\end{document}